\documentclass[12pt, hidelinks]{article}
\usepackage[small, bf]{caption}
\usepackage[utf8]{inputenc}
\usepackage{fullpage}
\usepackage{appendix}
\usepackage{graphicx}
\usepackage{amsmath}
\usepackage{amssymb}
\usepackage{amsthm, amsfonts}
\usepackage{url}

\newcommand{\downto}{\downarrow}

\newcommand{\eps}{\varepsilon}

\newcommand{\reals}{{\mbox{\bf R}}}

\newcommand{\eg}{{\it e.g.}}
\newcommand{\ie}{{\it i.e.}}

\newcommand{\BEAS}{\begin{eqnarray*}}
\newcommand{\EEAS}{\end{eqnarray*}}
\newcommand{\BEA}{\begin{eqnarray}}
\newcommand{\EEA}{\end{eqnarray}}
\newcommand{\BEQ}{\begin{equation}}
\newcommand{\EEQ}{\end{equation}}
\newcommand{\BIT}{\begin{itemize}}
\newcommand{\EIT}{\end{itemize}}

\title{Replicating Monotonic Payoffs Without Oracles}

\author{
Guillermo Angeris\\
{\small \texttt{angeris@stanford.edu}}
\and
Alex Evans\\
{\small \texttt{ahe4nc@gmail.com}}
\and
Tarun Chitra\\
{\small \texttt{tarun@gauntlet.network}}}

\date{September 2021}

\begin{document}

\maketitle

\begin{abstract}
In this paper, we show that any monotonic payoff can be replicated using only 
liquidity provider shares in constant function market makers (CFMMs),
without the need for additional collateral or oracles. Such payoffs include cash-or-nothing calls 
and capped calls, among many others, and we give
an explicit method for finding a trading function matching these payoffs.
For example, this method provides an easy way to show that the trading function for
maintaining a portfolio where 50\% of the portfolio is allocated in one asset and 50\% in the 
other is exactly the constant product market maker (\eg, Uniswap) from first principles. We 
additionally provide a simple
formula for the total earnings of an arbitrageur who is arbitraging against these CFMMs.

\end{abstract}

\section*{Introduction}
Constant function market makers (CFMMs) continue to be the most used decentralized application, with billions of dollars in daily trading volume.
Recently, there has been increased interest in understanding the space of financial products that can be created with CFMMs.
The main usage of CFMMs, to date, is as decentralized exchanges, or DEXs for short.
When CFMMs are used as DEXs, liquidity providers pool assets into the CFMM contracts with the aim of earning passive yield on their assets,
while traders, who wish to swap one asset for another, execute trades against this pool while paying a small fee to the liquidity providers on each trade.
To decide which trades to accept, the CFMM contract ensures that a specific function, known as the \emph{trading function},
that depends on the current state of the contract and the traded amounts, is kept constant.
An important consequence of this design is that arbitrageurs will always ensure that the price quoted by CFMM stays synchronized with the price of other markets that consist of the
same asset pairs~\cite{angeris2019analysis, angeris2020improved}. 

\paragraph{Previous work.}
Another view of a liquidity provider's shares in a CFMM is as a dynamically adjusted portfolio held by liquidity providers. This portfolio is constantly adjusted by 
arbitrageurs~\cite{evansOptimalFees} as the price of assets in external markets changes.
It has been shown that these portfolios held by liquidity providers can replicate a number of other financial payoffs~\cite{clark2020replicating, rmms, clark2021replicating}.
Evans~\cite{evans2020liquidity} was the first to show that CFMMs with dynamic trading functions (\ie, where the trading function is allowed to evolve over time) can replicate
\emph{any} unlevered payoff function.
If leverage is available (\eg, if users are allowed to borrow assets against a liquidity provider's share of the pool) it is possible to replicate a number of bounded convex
payoffs~\cite{chitra2021note}.
Furthermore,~\cite{rmms} provides a simple method to construct a CFMM trading function given a desired portfolio value function for liquidity providers. 
This method depends on the concavity of the payoff function and uses some basic properties of convex functions to construct the desired payoffs.
However, the payoffs that can be generated via this method are restricted --- one can only construct a CFMM whose liquidity provider shares have a portfolio value function that
is concave, nonnegative, nondecreasing, and 1-homogeneous, the so-called `consistent' portfolio values. In fact, the paper shows that the space of CFMMs and the space of consistent
portfolio value functions are equivalent in that every CFMM has a consistent portfolio value function and vice-versa.

\paragraph{Comparisons and this work.}
The key insight of this work is to note that liquidity providers do not have to sell their entire position, but can instead sell off the right to
withdraw individual components of the share. Surprisingly, this small fact allows a much richer set of payoffs to be replicated without leverage,
without additional collateral than what is held in a CFMM liquidity provider share, and without requiring an external oracle~\cite{eskandari2021sok}.
More specifically, we show that any monotonic payoff, in a certain price interval, can be replicated using this technique by constructing an appropriate trading
function. In many practical cases, we show that these payoffs also yield explicit formulas for the trading function, and, in some special cases, the total expected arbitrage profit.

\paragraph{Replicating payoffs.}
Using this method, many of the capped or bounded convex payoffs of \cite{chitra2021note} can be replicated without requiring a CFMM lender.
Specifically, we show that a number of payoffs such as capped calls, logarithmic payoffs, and the cash-or-nothing call (sometimes called the binary option) can be easily replicated.
We also show that one can uniquely derive the most popular trading function, the constant mean market maker~\cite{uniswap, adamsUniswapV2, balancer, evans2020liquidity}
of Uniswap and Balancer, by requiring that the user's portfolio have a constant proportion of their total portfolio value in each asset.
This illustrates that liquidity providers should be assessing the quality of a CFMM pool by assessing the associated payoffs, rather than just the trading function itself.
Provided that a user can construct smart contracts that securitize a subset of the components of the portfolio in the liquidity provider share, it is possible to replicate these
payoffs. We note that there are live smart contracts on Ethereum, such as Ondo Finance~\cite{ondo_vault} that allow for liquidity
providers to sell a portion of their yield to other market participants.

\paragraph{Summary.}
In~\S\ref{sec:replicating-payoffs}, we present all of the mathematical background needed to construct the trading function for a given payoff, conditions
under which payoffs that span an infinite interval can be replicated, and a simple formula for the earnings of an arbitrageur for a given price process. We
then provide examples of different payoffs that are useful in practice in~\S\ref{sec:examples} and discuss some interpretations
and extensions in~\S\ref{sec:further-discussion}.

\section{Replicating payoffs}\label{sec:replicating-payoffs}
A \emph{payoff} is a function $f: \reals_+ \to \reals$ which maps the \emph{price} $p \ge 0$ of some asset, which we will
call the risky asset, to
a quantity of the num\'eraire, $f(p)$. A \emph{monotonic payoff} is one where the payoff function $f$ is monotonically nondecreasing; \ie,
\[
f(p) \le f(q),
\]
whenever $p \le q$. In the special case that $f$ is differentiable, this is equivalent to $f'(p) \ge 0$ for all $p \ge 0$. Throughout the rest of this
paper, we will be concerned only with the monotonic payoffs that are nonnegative (\ie, satisfy 
$f(p) \ge 0$) but show some extensions which relax this latter condition. Additionally, for 
convenience, we will assume
that $f$ is differentiable for the rest of this section, but we will present results
that apply in much more general settings such as
when $f$ is differentiable, except at a countable number of discontinuities, and is lower 
semicontinuous.

\paragraph{Portfolio interpretation.} We can view the monotonic payoff as a user's desired portfolio allocation over two assets: the risky asset and the
num\'eraire, with $f(p)$ denoting how much of the num\'eraire the user is holding at price $p$. In this scenario, the user is continually rebalancing the portfolio (\eg, by selling or buying the risky asset) as the
market price changes. Using this strategy, if the market price changes from $p_1 > 0$ to $p_2 > 0$, the total amount of risky asset that the user needs
to sell is given by:
\begin{equation}\label{eq:risky-amount}
\int_{p_1}^{p_2} \frac{f'(p)}{p}\,dp.
\end{equation}
A basic argument for this is as follows.
Note that, as the price changes from $p$ to $p + h$, the user must buy $f(p+h)-f(p)$ of the num\'eraire (selling if the quantity is
negative). If $h$ is small, this costs approximately $(f(p + h) - f(p))/p$, and, since $f$ is differentiable, we have:
\[
\frac{f(p + h) - f(p)}{p} = \frac{f'(p)h}{p} + o(h).
\]
Dividing the right hand side by $h$, taking the limit, and integrating over $p$ gives the desired result. This argument can be made fully
airtight, but we omit the details as they are mostly mechanical and unenlightening.

%
\paragraph{Required risky asset.} A natural question is: at some price $\alpha \le p \le 
\beta$ where $\alpha$, $\beta$ satisfy $0 \le \alpha \le \beta \le \infty$ (we allow that $\beta = \infty$ for convenience), how much of the risky
asset do users need to hold in their portfolios to be able to trade using this strategy? This follows 
from~\eqref{eq:risky-amount}, which gives that
\[
g(p) = \int_p^\beta \frac{f'(q)}{q}\,dq.
\]
Therefore, at any price $\alpha \le p \le \beta$, the portfolio under this strategy contains $f(p)$ of the num\'eraire, and $g(p)$ of the risky asset. Additionally,
we note that, since $f' \ge 0$, the function $g$ is nonincreasing and nonnegative.
We call $g(p)$ the `replication cost' at price $p$.

\subsection{Desired portfolios and trading functions}\label{sec:desired-portfolios}
We will define the set of \emph{desired portfolios} as:
\begin{equation}\label{eq:portfolios}
S = \{(f(p), g(p))\mid \alpha \le p \le \beta\}.
\end{equation}
In other words, the set $S$ is the set of all possible portfolio allocations within the replication interval $[\alpha, \beta]$.

\paragraph{Portfolio value.} A natural question to ask is: what is the portfolio value of $S$, at some price $p$?
The portfolio value $V(p)$ at price $p$ is defined as the sum of the total value of all assets, in terms of the num\'eraire, \ie:
\begin{equation}\label{eq:portfolio-value}
V(p) = f(p) + pg(p).
\end{equation}
We can also rewrite $V$ in an equivalent, but very useful way:
\begin{equation}\label{eq:pv-rewrite}
    V(p) = V(\alpha) + \int_{\alpha}^p g(q)\,dq.
\end{equation}
To see this, note that:
\[
\int_{\alpha}^p g(q)\,dq = pg(p) - \alpha g(\alpha) - \int_\alpha^p q\,dg(q),
\]
where the right hand side follows from integration by parts. Since the last term of
this expression is equal to:
\[
-\int_\alpha^p q\,dg(q) = \int_\alpha^p q\frac{df(q)}{q} = f(p) - f(\alpha),
\]
then~\eqref{eq:pv-rewrite} follows from a basic rearrangement.

The portfolio value function has several immediate properties. For example, 
we can see from~\eqref{eq:portfolio-value} it is nonnegative 
since $f(p), g(p) \ge 0$ whenever $0 \le \alpha \le p \le \beta$.
On the other hand, using~\eqref{eq:pv-rewrite}, we can see that
$V$ nondecreasing since $g \ge 0$, and concave since $g$ is nonincreasing.
Putting this all together, we see that the portfolio value $V$ is a 
nonnegative, nondecreasing, concave function, which implies that
there is a constant function market maker with portfolio value function
$V$~\cite{rmms}.

\paragraph{Trading function.} We can construct a trading function
for a constant function market maker whose liquidity provider position
at price $p$ is equal to $(f(p), g(p))$. (For a general introduction to
CFMMs and trading functions, see, \eg,~\cite[\S2]{angeris2021constant}.)
To do this, we use
a simplification of the results from~\cite{rmms}, given in appendix~\ref{app:rmms}.
This simplification shows that, given the nonnegative, nondecreasing,
concave portfolio value function $V$, we can write a trading function
\begin{equation}\label{eq:trading-inf}
\psi(R_1, R_2) = \inf_{\alpha \le p \le \beta} \left(R_1 + pR_2 - V(p)\right),
\end{equation}
whose portfolio value is equal to $V(p)$ at price $p$, for all $R_1, R_2 \ge 0$. Using~\eqref{eq:pv-rewrite},
then
\[
pR_2 - V(p) = \alpha R_2 - V(\alpha) + \int_\alpha^p (R_2 - g(p))\,dp.
\]
Since $g$ is nonincreasing, then it is immediate that this expression is minimized by choosing
$p$ to be equal to
\[
g^{-1}(R_2) = \sup\,\{\alpha \le p \le \beta \mid g(p) \ge R_2\},
\]
if the set is nonempty, and we set $g^{-1}(R_2) = \alpha$, otherwise. To see this, note that the
integrand of the term
\[
\int_\alpha^p (R_2 - g(p))\,dp
\]
is nonpositive for all $p \le q$ where $q$ satisfies $g(q) \ge R_2$, and choosing
the largest such $q$, \ie, $q = g^{-1}(R_2)$, as defined previously, minimizes the total 
integral.
We use the suggestive `inverse' notation since $g^{-1}$ is indeed the inverse of $g$ when $g$ is
continuous and $R_2$ satisfies $g(\beta) \le R_2 \le g(\alpha)$, which is a
common case in practice.

Plugging this back into~\eqref{eq:trading-inf}, the trading function is given by
\begin{equation}\label{eq:trading-function}
\psi(R_1, R_2) = R_1 + g^{-1}(R_2)R_2 - V(g^{-1}(R_2)).
\end{equation}
In the common special case that $g(g^{-1}(R_2)) = R_2$ (\ie, $g^{-1}(R_2)$ is the `true inverse') then this simplifies to:
\begin{equation}\label{eq:trading-simple}
\psi(R_1, R_2) = R_1 - f(g^{-1}(R_2)),
\end{equation}
which, from the previous discussion happens when, \eg, $g$ is continuous.

\paragraph{Discussion.} The function $\psi$ in~\eqref{eq:trading-inf} is increasing and it is concave as it is the infimum over a family of functions
that are linear in $R_1$ and $R_2$~\cite[\S3.2.3]{cvxbook}. Because of this, arbitrageurs are incentivized to arbitrage against
a CFMM using $\psi$ as its trading function, in order to make its price match that of an external market. From the previous discussion, this ensures that the liquidity provider's
portfolio, when the external market is at price $p$, is given by $(f(p), g(p))$. We will show a direct proof of this in~\S\ref{sec:arb-earning} and give a simple expression for the total earnings of arbitrageurs who arbitrage
against CFMMs of this form.

There are a few possible implementations of this system, and we present a very simple one.
In this implementation, a liquidity provider first mints a share on a CFMM whose trading
function is specified above by putting in $(f(p), g(p))$ of
the num\'eraire and risky assets, respectively, when the market price is $p$. The
liquidity provider then sells off the right to only the num\'eraire
side of the pool, which has value $f(p)$, via some token. This token grants the buyer the following
right: when the token is burned, the liquidity provider share is also burned, and the amount
of num\'eraire found in the pool is immediately paid out to the token holder,
while the remaining risky asset is paid out to the liquidity provider.
The underlying value of the token is at least (but often close to) $f(p)$, as it can always be redeemed for that amount at any point,
which is the desired monotonic payoff of the price $p$.

\subsection{Replication costs}
In the case that $\beta < \infty$ and $p > 0$, the amount of risky asset needed to replicate
the portfolio is
always finite when $f$ is finite, but this need not be true as $\beta \to \infty$. We show
that the growth of $f$ needs to be at most sublinear in order to be finite, while it suffices to 
be polynomial of degree $(1-\eps)$ to be finite, for any $p > 0$.

\paragraph{Necessary conditions.} The payoff function $f$ has to exhibit sublinear growth in order to have a finite replication cost. More specifically, if $f$ is not sublinear in that
\[
f(p) \ge Cp,
\]
for some $C > 0$ and for all $p \ge p_0 \ge \beta$ (\ie, if $f(p) = \Omega(p)$)
then, for any $p$
\[
g(p) = \int_{p}^\beta \frac{f'(p)}{p}\,dp \ge \int_{p_0}^\beta \frac{f'(p)}{p}\,dp \ge \int_{p_0}^\beta \frac{C}{p}\,dp \to \infty,
\]
as $\beta \to \infty$. So it is necessary that $f$ grows slower than any linear function 
in order for the risky side of the portfolio to be finite for any $p$. In other words, we must 
have that $f$ is little-oh of $p$, \ie, $f(p) = o(p)$ as $p\to \infty$. This means that, \eg,
it is not possible to replicate a call option with a finite amount of risky asset.

\paragraph{Sufficient condition.} In order to have finite replication cost for all prices $p > 0$, it suffices for the payoff function $f$ to have $\alpha > 0$ and a growth of at most
\[
f(p) \le C p^{1-\eps},
\]
for some $C > 0$, $\eps > 0$ and for all $p \ge p_0$. The proof is nearly identical to the previous, since, for any $p > 0$,
\[
g(p) \le \int_p^{p_0} \frac{f'(p)}{p}\,dp + \int_{p_0}^{\beta} \frac{f'(p)}{p}\,dp \le D + C(1-\eps)\int_{p_0}^\beta p^{-(1+\eps)}\,dp \to  D + C(1-\eps)\frac{1}{p_0^\eps\eps},
\]
where $D = \int_p^{p_0}f'(p)/p\,dp$ is finite since $p > 0$, and we assume $\eps < 1$ for convenience. (If $\eps \ge 1$ in the original statement, then we can always choose $\eps < 1$ as it is a strictly worse bound.)

\subsection{Arbitrageur earnings}\label{sec:arb-earning}
From~\S\ref{sec:desired-portfolios}, we know that, because the portfolio given by $(f(p), g(p))$ yields a concave payoff, then arbitrageurs will be incentivized to arbitrage against it
and will make the price of the CFMM match that of an external market (see, \eg,~\cite{angeris2020improved}). We prove this directly here,
using a slightly different method.

More specifically, given that the arbitrageur is able to choose any $q$ which leads to a portfolio allocation of $(f(q), g(q))$ for the liquidity provider,
we show that, if an external market has price $\alpha \le p \le \beta$, then $p$ is an optimal choice for the portfolio allocation in that it maximizes the
arbitrageur's earnings.

\paragraph{Maximum profit.} Arbitrageurs seek to maximize their own profit and therefore minimize the portfolio value of the
liquidity provider. This is expressed by the following problem:
\begin{equation}\label{eq:min-lp}
\begin{aligned}
    & \text{minimize} && f(q) + pg(q),
\end{aligned}
\end{equation}
with variable $\alpha \le q \le \beta$. Taking the derivative of the objective and using the definition of $g$, this can be written
\[
f'(q) + pg'(q) = f'(q) - p \frac{f'(q)}{q} = f'(q)\left(1 - \frac{p}{q}\right).
\]
Note that, since $f' \ge 0$ by definition, then the objective is unimodal as its derivative changes signs at most once.
Additionally, the first order optimality condition,
\[
f'(q)\left(1 - \frac{p}{q}\right) = 0,
\]
implies that choosing $p/q = 1$ (that is, $p=q$) is a solution with portfolio value $f(p) + pg(p)$, when the market price
is $p$.

\paragraph{Nonnegativity of profit.} The profit of an arbitrageur is the negative of the change in the portfolio value
when the price changes from $p$ to $p'$, which is
\begin{equation}\label{eq:arb-payoff}
p'(g(p) - g(p')) + f(p) - f(p').
\end{equation}
Because $p$ is a feasible point for the original problem~\eqref{eq:min-lp}, then we have that
\[
f(p') + p'g(p') \le f(p) + p'g(p),
\]
which yields, after rearrangement:
\[
p'(g(p) - g(p')) + f(p) - f(p') \ge 0.
\]

\paragraph{Total earnings.}
Given a sequence of price changes $p_0, p_1, \dots, p_n$, the total earnings, using~\eqref{eq:arb-payoff}, are
\[
\sum_{i=1}^n p_i(g(p_{i-1})-g(p_{i})) + f(p_{i-1}) - f(p_{i}) = f(p_0) - f(p_n) - \sum_{i=1}^n p_i(g(p_{i})-g(p_{i-1})).
\]
Taking limits, we find that the total earnings are
\[
W = f(P_0) - f(P_T) - \int_0^T P_t\,dg(P_t),
\]
where $P_t$ is a price process with $0 \le t \le T$, and the integral is to be interpreted in the It\^o sense. Integrating by parts, we can write this as the slightly more interpretable form:
\begin{equation}\label{eq:arb-earnings}
W = V(P_0) - V(P_T) + \int_{0}^T g(P_t)\,dP_t,
\end{equation}
where $V$ is the portfolio value function described in~\eqref{eq:portfolio-value}.
From the previous discussion, we know that the total arbitrageur profit is nonnegative,
$W \ge 0$, for any price process $P_t$.

\paragraph{Discussion.} We can interpret the individual terms in the 
arbitrageur's total earnings, equation~\eqref{eq:arb-earnings}, in a simple way. The first term, 
$V(P_0) - V(P_T)$ is the negative of the payoff to the liquidity provider, which is received
from having the underlying portfolio rebalanced as the price of the asset changes from time
$t=0$ to time $t = T$. The second term, $\int_0^T g(P_t)\,dP_t$, is a `path-dependent' term
that comes from the arbitrageur performing arbitrage at each time period.

\section{Examples}\label{sec:examples}
In this section, we provide a few useful examples of payoffs which can easily be implemented
in practice using the methods presented in the previous section.

\subsection{Cash-or-nothing call}
The simplest monotonic function which can be replicated is perhaps the \emph{cash-or-nothing call}, which has payoff:
\[
f(p) = \begin{cases}
    0 & p \le p_0\\
    1 & p > p_0,
\end{cases}
\]
for some $p_0 > 0$ and all $p \ge 0$ (\ie, $\alpha = 0$ and $\beta = \infty$). The amount of risky asset is given by:
\[
g(p) = \int_p^\infty \frac{df(q)}{q} = \frac{1-f(p)}{p_0}.
\]
(This integral can be interpreted in many ways, as the derivative of $f$ doesn't exist at $p = p_0$,
with perhaps the simplest being a Riemann--Stieltjes integral since $f$ is monotonic and $1/p$ is differentiable.) 

\paragraph{Interpretation.} In fact, because the cash-or-nothing call is so simple, the functions 
$f$ and $g$ can be intuited and proven without needing the explicit formula given 
in~\eqref{eq:risky-amount}. Since every nonnegative, monotonic function can be written as the 
limit of the sum of a number of cash-or-nothing calls, we can recover~\eqref{eq:risky-amount} 
using only this reasoning. This provides an alternative method for recovering most of the results
provided in this paper and is very similar in spirit to the Carr--Madan
replication method in finance~\cite[App.~1]{carr2001towards}.

\paragraph{Trading function.}
Because $g$ is not continuous, there is no guarantee that there exists a $p \ge 0$ such that
$g(p) = R_2$, even when $0 \le R_2 \le 1/p_0$, so the simplification in~\eqref{eq:trading-simple}
does not apply. On the other hand, it is not hard to show that
\[
g^{-1}(R_2) = \begin{cases}
p_0 & R_2 > 0\\
+\infty & \text{otherwise},
\end{cases}
\]
and therefore that $V(g^{-1}(R_2)) = 1$ for any $R_2 \ge 0$. The resulting trading function is
then
\[
\psi(R_1, R_2) = R_1 + p_0R_2 - 1,
\]
where we have defined $0\cdot \infty = 0$ for convenience. Note that this is just the
linear market maker, defined in~\cite[\S2.4]{angeris2021constant}.

\subsection{Capped call}\label{sec:clipped-call}
Another example of a payoff is that of the \emph{capped call}, whose payoff function is 
defined as 
\[
f(p) = \begin{cases}
    0 & p \le p_0 \\
    p - p_0 & p_0 < p \le p_1\\
    p_1 - p_0 & p > p_1,
\end{cases}
\]
where $0 < p_0 \le p_1$ are user-defined constants.
In this case, we have that
\[
g(p) = \begin{cases}
    \log(p_1/p_0) & p \le p_0\\
    \log(p_1/p) & p_0 < p \le p_1\\
    0 & p > p_1.
\end{cases}
\]

\paragraph{Trading function.} The trading function for a capped call can be easily computed
using the simplification provided in~\eqref{eq:trading-simple}, because the function $g$ is 
continuous. Using the definition of $g$, we have
\[
g^{-1}(x) = p_1e^{-x}
\]
whenever $0 \le x \le \log(p_1/p_0)$. The trading function can then be
written
\[
\psi(R_1, R_2) = R_1 - f(g^{-1}(R_2)) = R_1 + p_0 - p_1e^{-R_2},
\]
when $0 \le R_2 \le \log(p_1/p_0)$, and the reserves are otherwise invalid.

\subsection{Black-Scholes cash-or-nothing call}

The preceding examples give conditions for the terminal payoffs of binary and capped calls, respectively.
As discussed in~\cite{rmms}, it is often more useful to work with the price of a derivative contract under a parametric model. This typically requires significantly lower collateral to achieve a similar terminal payoff. One such model is that of Black-Scholes. Here we derive the payoff the a binary call under this model. In this case the payoff is defined as
\[
f(p)=\Phi(d(p))
\]
where $\Phi(\cdot)$ is the normal CDF and 
\[
d(p)=\frac{\log(p/K)-\tau\sigma^2/2}{\sigma \sqrt{\tau}}.
\]
Here, $\tau>0$ is the time to maturity, $K \geq 0$ is the strike price and $\sigma \ge 0$ is the implied volatility (we assume zero interest rates to minimize notation).

\paragraph{Integral simplification.} There is a simple but very useful `trick' to evaluate payoffs of the form
\[
f(p) = r(s(p)),
\]
where $s$ is a monotonically increasing function. Given that we want to evaluate $g$, then:
\[
g(p) =\int_p^\infty \frac{r'(s(q))s'(q)}{q}\,dq = \int_{s(p)}^{s(\infty)} \frac{r'(u)}{s^{-1}(u)}\,du,
\]
where the last equality follows by using the $u$-substitution $u = s(q)$.

\paragraph{Replication cost.} Using this trick, we have that the amount of risky asset is given by:
\[
g(p) = \frac{1}{K} \int_{d(p)}^\infty \frac{\phi(u)}{\exp(\sigma\sqrt{\tau}u+ \tau\sigma^2/2)}du = \frac{1}{K}\left(1-\Phi(d(p)+\sigma \sqrt{\tau})\right),
\]
where $\phi(q) = \Phi'(q) = \exp(-q^2/2)/\sqrt{2\pi}$.
As before, we may use~\eqref{eq:trading-simple}, because the function $g$ is continuous. Using the definition of $g$, we have
\[
g^{-1}(x) = Ke^{\sigma \sqrt{\tau}\Phi^{-1}(1-Kx)-\tau\sigma^2/2}
\]
From~\eqref{eq:trading-simple}, after some cancellations we have
\[
\psi(R_1, R_2) = R_1 - \Phi(\Phi^{-1}(1-KR_2)-\sigma \sqrt{\tau})
\]
which coincides with the trading function derived in~\cite{rmms} for a covered call under Black-Scholes. (As noted in~\cite{rmms}, replicating in this manner requires additional capital
as the portfolio value function is pointwise strictly decreasing in the time to maturity, $\tau$. This `gain' as $\tau \downto 0$ is sometimes called ``theta" in finance.)

\subsection{Logarithmic payoff}
It is also possible to replicate more complicated payoffs. One such example is the 
\emph{logarithmic payoff} which is given by
\[
f(p) = \begin{cases}
    0 & p < p_0\\
    \log(p/p_0) & p \ge p_0.
\end{cases}
\]
Using the above, then the replication cost at $p$ is:
\[
g(p) = \begin{cases}
    1/p_0 & p < p_0\\
    1/p & p \ge p_0.
\end{cases}
\]
Note that the total amount to cover goes to infinity as $p_0 \downto 0$.

\paragraph{Trading function.} The trading function is also an exercise in algebra, since
the function $g$ is continuous and we can use~\eqref{eq:trading-simple}:
\[
\psi(R_1, R_2) = R_1 + \log(p_0R_2),
\]
where the range of valid reserves is $0 \le R_2 \le 1/p_0$.

\paragraph{Arbitrageur earnings.} Using equation~\eqref{eq:arb-earnings}, we can find how much
an arbitrageur should expect to earn when the price of the risky asset follows a geometric Brownian motion with stochastic differential 
\[
dP_t = P_t\sigma dW_t,
\]
where $W_t$ is a standard Brownian motion. The true portfolio value function is given by
\[
V(p) = \begin{cases}
p/p_0 & p < p_0\\
1 + \log(p/p_0) & p \ge p_0.
\end{cases}
\]
This portfolio value function is difficult to handle directly, so we will consider the following portfolio value function:
\[
V(p) = 1+\log(p/p_0),
\]
which is `approximately' equal to the true portfolio value function when $p_0$ is small. Using this latter approximation, we then have:
\[
W = \log[P_0/P_T]+\int_0^T \sigma dW_t.
\]
Taking expectations gives
\[
E[W] = \frac{1}{2} \sigma^2 T.
\]
In other words, the arbitrageur's expected payoff (approximately) matches that of a variance swap~\cite{Neuberger74}.
Loosely speaking, the right to arbitrage a no-fee CFMM with a logarithmic payoff may be used to replicate a variance swap without an oracle.
Of course, in practice, the above argument is impractical as it requires unbounded capital (as $p_0 \downto 0$ is required),
but we expect that finite reserve amounts will suffice to provide similar payoffs.

\subsection{Capped power payoffs}
Given any power, we can find the market maker for the payoff
\[
f(p) = \begin{cases}
0 & p < p_0\\
p^\alpha - p_0^\alpha & p_0 \le p \le p_1\\
p_1^\alpha - p_0^\alpha & p > p_1,
\end{cases}
\]
where $0 \le p_0 \le p_1 \le \infty$ and $\alpha \in \reals$ is a real number.
It is an exercise in integration to show
\[
g(p) = \frac{\alpha}{\alpha - 1} (p_1^{\alpha-1}- p^{\alpha-1}),
\]
if $p_0 \le p \le p_1$, while $g(p) = g(p_0)$ if $p < p_0$ and $g(p) = 0$, otherwise.
Note that $p_1 < \infty$ is necessary if $\alpha \ge 1$ in order to have a finite replication
cost. This portfolio generalizes that of the the capped call presented
in~\S\ref{sec:clipped-call} as the special case when $\alpha = 1$.

\paragraph{Trading function.} We can write
\[
g^{-1}(x) = \left(p_1^{\alpha - 1} + \frac{1-\alpha}{\alpha}x\right)^{1/(\alpha-1)},
\]
with domain $0 \le x \le g(p_0)$. Because $g$ is continuous, we can use the
simplification in~\eqref{eq:trading-simple}, so the trading function can be written as:
\[
\psi(R_1, R_2) = R_1 + p_0^{\alpha} - \left(p_1^{\alpha-1} + \frac{1-\alpha}{\alpha}R_2\right)^{\alpha/(\alpha-1)},
\]
where the reserves are valid only when $0 \le R_2 \le g(p_0)$.

\subsection{Constant proportion portfolios}
What happens if a liquidity provider is seeking a constant proportion of their wealth
to be in one asset or the other? As is well known from the literature, this is done
by the class of constant mean market makers~\cite{balancer, adamsUniswapV2}, but we can also
easily recover this from first principles using this framework.

\paragraph{Requirements.} In this problem, a user wishes to have a portfolio allocation
$(f(p), g(p))$ of the num\'eraire and risky assets at each price $p$ such that $0 < w < 1$ of
the portfolio value is in the num\'eraire, while $(1-w)$ of the portfolio value is
in the risky asset; in other words, $f$ and $g$ must satisfy
\[
wf(p) = (1-w)pg(p).
\]
Dividing by $p$, taking the derivative of both sides, and using the definition of $g$, we have:
\[
pf'(p) = wf(p),
\]
after some simplifications. It is a standard exercise to show that the unique family
of solutions $f$ to this problem is given by
\[
f(p) = Cp^{w},
\]
where $C \ge 0$ is a constant that determines the total portfolio value. This implies
that the function $g$ is given by:
\[
g(p) = C\frac{w}{1-w}\frac{1}{p^{1-w}}. 
\]
Note that $g$ is a continuous function on the positive reals.

\paragraph{Trading function.} Using the definition of $g$ above, we can write
\[
g^{-1}(x) = \left(\frac{1-w}{w}\frac{x}{C}\right)^{-\frac{1}{1-w}},
\]
which implies
\[
\psi(R_1, R_2) = R_1 - f(g^{-1}(R_2)) = R_1 - \left(\frac{R_2}{C}\frac{1-w}{w}\right)^{-\frac{w}{1-w}}.
\]
By multiplying both sides of the equation by the right-most term and taking the $(1-w)$th power,
we can write the equivalent trading function:
\[
\tilde\psi(R_1, R_2) = R_1^{1-w}R_2^w - C',
\]
where $C' = (Cw/(1-w))^{w/(1-w)}$. The trading functions are equivalent in the sense that $R_1$ 
and $R_2$ satisfy $\psi(R_1, R_2) \ge 0$ if, and only if, $\tilde \psi(R_1, R_2) \ge 0$. This 
is the classic trading function for constant mean market makers.

\section{Further discussion}\label{sec:further-discussion}
In this section, we share some further thoughts and (nearly immediate) extensions of the method presented above.

\paragraph{Staking.}
Proof of Stake (PoS) protocols allow a user to lock a digital asset into a pool (a process known as \emph{staking}) in order to provide a service to the network.
In return, the network distributes block rewards (a form of subsidy) and collected fees to stakers.
The most popular usage of PoS networks is executing consensus algorithms of replicating state machines, such as blockchains.
Many CFMMs exist as contracts whose state updates and execution are maintained by a decentralized PoS network.

A number of protocols such as Osmosis \cite{agrawal_osmosis} and Penumbra \cite{zswap} allow users to stake not only the underlying staking asset but
the liquidity provider shares that contain the staking asset to receive consensus rewards.
For instance, if a network uses a risky asset A for staking, such networks would allow users to stake any A-num\'eraire LP share to receive rewards.
One reason a network might want to incentivize such a pool is to ensure that there is sufficient on-chain liquidity to purchase asset A with the num\'eraire, regardless of external market liquidity. 
The rewards earned by liquidity provider share stakers would be lower than those earned by those staking the base asset.
However, liquidity providers would effectively be compounding their returns as they get both CFMM fees and staking rewards.

A PoS system with liquidity provider share staking also can be used to construct staking derivatives, which allows for improved capital efficiency.
This capital efficiency comes from allowing liquidity providers to borrow against their locked shares, provided that price of the underlying is above a certain threshold. 
Staking derivatives can be constructed as capped monotone payoffs~\cite{chitra2020stake} and the construction of monotone replication in \S1.1 can be enforced by a consensus protocol.
In particular, while the right to execute the option to liquidate an LP share and realize a position of $f(p)$ unit of num\'eraire is only bounded below by $f(p)$ in general, it can be made to exactly equal $f(p)$ when a consensus protocol buys the right to execute an LP share liquidation.
This is because a consensus protocol controls when the precise liquidation occurs and if added as a consensus rule, will only happen \emph{exactly} when the portfolio value dropped below a threshold.
This suggests that staking derivatives are natural consumers of monotone replication.

\paragraph{Continuous liquidation interpretation.}
Another interpretation of monotone replication via a token that allows for one side of the LP share to be liquidated is as a loan that is being continuously liquidated as the price
changes. More specifically, suppose that an LP share is created with $(f(p_0), g(p_0))$ units of the risky asset and num\'eraire, respectively.
Further, suppose that rights to the num\'eraire portion of the pool are sold (as in~\S\ref{sec:desired-portfolios}).
Now, if the price changes from $p_0$ to $p_1$ over a time interval of length $h$, then
holding share through the price change is the same as selling the option for $g(p_0)$
units of num\'eraire, having it executed at $p_1$, and using the proceeds to mint a new share which has the allocation $(f(p_1), g(p_1))$.
If we take $h \rightarrow 0$, then this equivalence can be viewed as saying that holding the liquidity provider share is equivalent to continuously
rebalancing by selling num\'eraire rights and using the proceeds upon execution to recreate another share and resell num\'eraire rights.

\paragraph{Negative reserves.} Note that we do not require that $f$ be nonnegative at any
point in the presentation except to assert that the portfolio value function is nonnegative.
We can, instead, allow $f(p)$ to be negative in so far as the portfolio value function is 
nonnegative at all times.
This idea corresponds to the fact that the portfolio is allowed to short an asset, 
so long as the position remains solvent, \ie, so long as the portfolio
value function is nonnegative. We can relax this further by requiring only that
the portfolio position is nonnegative in an interval of prices, while making sure that the 
portfolio holder's collateral is liquidated when the price leaves this interval.

\section{Conclusion}
We have shown how to replicate nonnegative, monotonic payoff functions (somewhat generalizing the result of~\cite{rmms}) by allowing liquidity providers to sell
the rights to a specific component of their portfolio, rather than their entire portfolio.
This generalization allows us to realize a number of payoffs that were previously thought to require explicit leverage to replicate with CFMM liquidity provider shares.
Such a replication increases the space of unlevered structured products that can be created with liquidity provider shares which don't require additional collateral.
Our results are likely best implemented in proof-of-stake (PoS) systems that interact with liquidity provider shares to construct simple staking derivatives.
Here, the PoS network is the buyer of 
num\'eraire exposure. Possible future work directions include adding the effect of fees on monotone replications and computing expected liquidity provider profits in different fee regimes
after selling a component of the underlying portfolio. 

\bibliographystyle{alpha}
\bibliography{citations.bib}

\appendix

\section{RMMs simplification}\label{app:rmms}
In~\cite{rmms}, it is shown that, given a portfolio value function $U: \reals^2_+ \to \reals$ which is
nonnegative, nondecreasing, concave, and 1-homogeneous, the trading function defined as
\[
\psi(R_1, R_2) = \inf_{c_1, c_2} \left(c_1R_1 + c_2R_2 - U(c_1, c_2)\right)
\]
has $U$ as its portfolio value function. Additionally, given a scalar portfolio value function $V$,
that is nonnegative, nondecreasing, concave, and depends only on the relative price $p$ of
asset 1 to asset 2, it can be turned to a 1-homogeneous function $U$ that depends on the prices 
of both assets, by use of the perspective transform:
\[
U(c_1, c_2) = c_1 V(c_2/c_1)
\]
where we assume that $U(0, 0) = 0$ and $U(0, c_2) = -\infty$ for $c_2 > 0$. So, we can write:
\[
\psi(R_1, R_2) = \inf_{c_1, c_2} \left(c_1R_1 + c_2R_2 - c_1V(c_2/c_1)\right).
\]
Noting that 
\[
\psi(R_1, R_2) = \inf_{c_1}c_1 \left(\inf_{c_2} \left(R_1 + (c_2/c_1)R_2 - V(c_2/c_1)\right)\right)
\]
and letting $p = c_2/c_1$ we get
\[
\psi(R_1, R_2) = \inf_{c_1}c_1 \left(\inf_{p} \left(R_1 + pR_2 - V(p)\right)\right) = \begin{cases}
     0 & \inf_{p} \left(R_1 + pR_2 - V(p)\right) \ge 0\\
     -\infty & \text{otherwise}.
 \end{cases}
\]
It is nearly immediate that
\[
\tilde \psi(R_1, R_2) = \inf_{p} \left(R_1 + pR_2 - V(p)\right),
\]
is equivalent to $\psi$ in that $\psi(R_1, R_2) \ge 0$ if, and only if, 
$\tilde \psi(R_1, R_2) \ge 0$. The proof holds throughout even if $p$ is constrained to lie in the interval
$[\alpha, \beta]$ for $0 \le \alpha \le \beta \le +\infty$.

\end{document}